\documentstyle[12pt]{article}

\begin{document}
{\sf \begin{center} \noindent {\Large \bf Stretch-twist dynamo torus maps and unsteady flows in compact Riemannian magnetic flux tubes}\\[3mm]

by \\[0.3cm]

{\sl L.C. Garcia de Andrade}\\

\vspace{0.5cm} Departamento de F\'{\i}sica
Te\'orica -- IF -- Universidade do Estado do Rio de Janeiro-UERJ\\[-3mm]
Rua S\~ao Francisco Xavier, 524\\[-3mm]
Cep 20550-003, Maracan\~a, Rio de Janeiro, RJ, Brasil\\[-3mm]
Electronic mail address: garcia@dft.if.uerj.br\\[-3mm]
\vspace{2cm} {\bf Abstract}
\end{center}
\paragraph*{}
Earlier Arnold, Zeldovich, Ruzmaikin and Sokoloff [\textbf{JETP
(1982)}] have computed the eigenvalue of a uniform stretching torus
transformation which result on the first Riemann metric solution of
the dynamo action problem. Recently some other attempts to obtain
Riemann metrics representing dynamo action through conformal maps
have been undertaken [{\textbf{Phys. Plasmas 14 (2007)}]. Earlier,
Gilbert [\textbf{Proc. Roy. Soc. London A(2003)}] has investigated a
more realistic dynamo map solution than the one presented by Arnold
et al by producing a shearing of the Arnold's cat map, by eigenvalue
problem of a dynamo operator. In this paper, the eigenvalue of the
Riemann twisted torus dynamo flow metric is obtained as the ratio
between the poloidal and toroidal components of the flow. This
result is obtained from the Euler equation. In the twisted torus,
the eigenvalue of the Riemann metric is
${m}_{\pm}=\frac{1\pm{\sqrt{5}}}{2}$, which is quite close to the
value obtained by Arnold. In the case the viscosity Reynolds number
$Re\rightarrow{\infty}$, the torus flow is unstable as one
approaches the torus axis. In Arnold's dynamo metric the eigenvalues
are ${\chi}_{\pm}=\frac{3\pm{\sqrt{5}}}{2}$ which are very close to
the above value. Eigenvalues determine the growth rates of the
velocity ratio between poloidal and toroidal components of the flow.
The curved flow in torus follow previous work by Chen et al
[\textbf{Phys Fluids (2006)}]. The ${\alpha}$-effect dynamo is shown
to be a second-order effect in the torus curvature and velocity
flow. Loop dynamo flows and maps are also discussed.{\bf PACS
numbers:\hfill\parbox[t]{13.5cm}{02.40.Hw:differential
geometries.91.25.Cw-dynamo theories.}}}

\newpage
\newpage
 \section{Introduction}The investigation of rotating torus flows and their perturbation \cite{1} have been recently linked to the dynamo experiments
 with liquid sodium and other liquid metals \cite{2,3}. The incompressible flows, where most of
 the anti-dynamo theorems are applied \cite{4}, can be used as
 rotational flows in Perm torus \cite{2} and Riga experiment \cite{3}. Therefore a good
 understanding of the behaviour of these rotating torus flows is of
 utmost importance in planning these experiments. Alternative dynamo experiments using
 plasma flows have been recently obtained by Wang and his group
 \cite{5}, which has developed the first flowing magnetic plasma (FMP) experiment, called
 $P-24$ in order to detect dynamo action. In this paper a
 slight variation of the usual Riemann-flat (Riemann curvature zero) torus named a
 twisted torus \cite{6} filled with a rotating incompressible rotating flow. The incompressibility of the flow is well suited for liquid metals.
 In this twisted torus, as shown here, the Riemann curvature tensor does not vanishes and for thick tubes the torus is also stretched, a
 fundamental property for existence of dynamo action \cite{7}. This kind of twisted geometry is very usual in plasma devices such as heliotrons,
 stellarators and in the astrophysical and solar flux tubes \cite{8}. Previously Mikhailovskii \cite{9}, have made use of non-diagonal
 Riemann metrics to describe such a plasma devices. Here one follows Ricca \cite{8} approach and use a diagonal simpler twisted magnetic flux tube Riemann
 metric. The first example of a chaotic fast dynamo solution was found by Arnold et al \cite{10} by making
 use of a compressed and stretched Riemannian metric of the dynamo
 flow. The domain of the flow is given by a compact Riemannian manifold represented as the product of a torus $\textbf{T}^{2}$ and the closed interval $[0,1]$ of
 $0\le{z}\le{1}$ \cite{11}. This results in the Arnold Riemann metric
\begin{equation}
ds^{2}=e^{-{\lambda}z}dp^{2}+e^{{\lambda}z}dq^{2}+dz^{2} \label{1}
\end{equation}
which represents the stretching and contraction in distinct
Euclidean directions p and q , respectively induced by the
eigenvalues ${\chi}_{\pm}=\frac{3\pm{\sqrt{5}}}{2}$ also
corresponding to magnetic eigenvectors. The stationary dynamo flow
considered by Arnold was given by the simple uniformly stretching
flow $\textbf{v}=(0,0,v)$ in $(p,q,z)$ coordinates. More recent
attempts to build a dynamo action by making use of compact
Riemannian geometry includes the case of the fast dynamo of Chiconne
and Latushkin \cite{12}, and the conformally stretched fast dynamo
by Garcia de Andrade \cite{13}. The rotating torus flow equations
are
\begin{equation}
d_{t}{\textbf{v}}=-2{\vec{\Omega}}{\times}\textbf{v}+{\nu}{\Delta}\textbf{v}-\frac{{\nabla}p}{{\rho}_{0}}
\label{2}
\end{equation}
where ${\Delta}:={\nabla}^{2}$. The incompressible flows are
described by the solenoidal vector field as
\begin{equation}
{\nabla}.\textbf{v}=0 \label{3}
\end{equation}
Equation (\ref{2}) is the Navier-Stokes equation for the rotating
flow inside the torus, while ${\nu}$ is the viscosity constant. Here
\begin{equation}
d_{t}={\partial}_{t}+\textbf{v}.{\nabla} \label{4}
\end{equation}
Note that the eigenvalue problem in this case is not easy to define
since even in the absence of pressure gradients the eigenvalue would
be zero and if one tries to use the ABC flow for example
\begin{equation} {\vec{\Omega}}={\lambda}\textbf{v}
\label{5}
\end{equation}
the first term on (\ref{2}) vanishes but yet the eigenvalue problem
is not helped much. In this paper however, one presents an
eigenvalue like problem that solves the differential Euler equation
(\ref{2}). The scalar proportionality between the poloidal and
toroidal flows modulo the eigenvalue m shows that this eigenvalue is
quite similar to the torus eigenvalue obtained by Arnold. Actually
from the mathematical point of view these transformations define an
automorphism on the torus of the type $\cal{A}:
\textbf{v}\rightarrow{{\textbf{v}}}$. This paper is organised as
follows. In section II the Euler equation free of viscosity in the
background of a twisted torus flow. In this case the radial pressure
is unstable closer to the internal torus axis, as happens in some
torus plasma instability associated to coronal mass ejection,
recently investigated by Toerok and Kliem \cite{14}. The
$\alpha-dynamo$ effect is also discussed in section III. Section IV
addresses the loop filaments dynamo torus twist maps, while
discussions and conclusions appear in section V.
\newpage
\section{Stretch-twist dynamo maps}
Following previous work by Childress and Bayly \cite{13}, Gilbert
\cite{15} developed applied their stretch-fold-shear (SFS) dynamo
map in Arnold's cat map, by performing matrix transformations to his
the expression of cat dynamo Jacobian matrix \cite{4}
\begin{equation}
\textbf{M}_{\textbf{cat}}=\pmatrix{2&1\cr1&1\cr}\qquad \label{6}
\end{equation}
corresponding to the map $\textbf{x}\rightarrow{\textbf{Mx}}$. these
cat maps with shear, built by Gilbert result on the class of
matrices in the form:
\begin{equation}
\textbf{M}_{\textbf{cat-shear}}=\pmatrix{1+K^{2}&K\cr{K}&1\cr}\qquad
\label{7}
\end{equation}
where K is an integer related to the eigenvalues. In this paper one
shows that the torus twist map \cite{6}
\begin{equation}
\textbf{M}_{\textbf{twist}}=\pmatrix{1&1\cr0&1\cr}\qquad \label{8}
\end{equation}
which is clearly distinct from the chaotic dynamo map matrix
(\ref{6}), can be associated to the Ricca's Riemann twisted magnetic
flux tube metric
\begin{equation}
dl^{2}= dr^{2}+r^{2}d{{\theta}_{R}}^{2}+K^{2}(r,s)ds^{2} \label{9}
\end{equation}
under the transformation
\begin{equation}
\textbf{M}_{\textbf{twist}}=\pmatrix{1&-{\tau}_{0}\cr0&K(s)\cr}\qquad
\label{10}
\end{equation}
which is obtained from the transformation $x={\theta}_{R}$ and
$y=s$, which from the twist transformation
${\theta}(s):={\theta}_{R}-\int{{\tau}(s)ds}$, results in matrix
(\ref{10}), as long as one considers the helical case where the
Frenet torsion ${\tau}(s)={\tau}_{0}=constant$. In the case of thin
twisted magnetic flux tube, $K=1$ and the equation simplifies to
\begin{equation}
\textbf{M}_{\textbf{thin-tube}}=\pmatrix{1&-{\tau}_{0}\cr0&1\cr}\qquad
\label{11}
\end{equation}
which is still closer with the twist torus map. If one normalizes
the torsion ${\tau}_{0}=-1$ the last matrix reduces exactly to twist
torus map matrix (\ref{8}). Of course the presence of stretching
factor $K$ in the Riemann metric can be considering as a constant
stretch which reduces the eigenvalue matrix to
\begin{equation}
\textbf{M}_{\textbf{twist}}=\pmatrix{1&-{\tau}_{0}\cr0&{K_{0}}\cr}\qquad
\label{12}
\end{equation}
Let $\textbf{b}$ is the transpose of an initially constant toroidal
magnetic field given by \begin{equation}
\textbf{b}_{\textbf{twist}}=\pmatrix{0&1\cr}\qquad \label{13}
\end{equation}
Within these motivations, in the next section, one shall follow a
less mathematical path to help experimental dynamo physicists not
very familiar heavy framework of dynamo theory, and compute the
solution of the Euler equation with rotation in the case of a
homothetic motion between the poloidal and toroidal flow velocities.
It is shown that the self-induction magnetic equations eigenvalues
imply stretching or contraction of the magnetic flows and growth or
decay of the toroidal field with respect of the poloidal field, in
either case dynamo action is present.
\section{Unsteady twisted torus dynamo flows}
 Let us impose the constraint $d_{t}\textbf{v}=0$ on the Euler equation
\begin{equation}
d_{t}{\textbf{v}}=-2{\vec{\Omega}}{\times}\textbf{v}\label{14}
\end{equation}
Solutions of these equations yields naturally the ABC flows above.
The Euler equation can now be written in the background of the above
twisted flux tube Riemann metric, where
${\theta}(s):={\theta}_{R}-\int{{\tau}(s)ds}$ and $r_{0}$ is the
constant radius of the constant cross-section flux tube, and
$K(s)=(1-r{\kappa}(s)cos{\theta}(s))$. If the tube is thin factor
$K(s)\approx{1}$ and the gradient operator is given by
\begin{equation}
{\nabla}=\textbf{t}{K}^{-1}{\partial}_{s}+\textbf{e}_{\theta}\frac{1}{r}{\partial}_{\theta}+\textbf{e}_{r}{\partial}_{r}\label{15}
\end{equation}
where the Riemannian line element is in general given by
\begin{equation}
dl^{2}= g_{ij}dx^{i}d{x}^{j} \label{16}
\end{equation}
where $(i,j=1,2,3)$ and $x^{j}\in{\textbf{R}^{3}}$. To express Euler
equations in this metric background one must use the dynamical
relations from vector analysis and the theory of curves in the
Frenet frame $(\textbf{t},\textbf{n},\textbf{b})$ are
\begin{equation}
\textbf{t}'=\kappa\textbf{n} \label{17}
\end{equation}
\begin{equation}
\textbf{n}'=-\kappa\textbf{t}+ {\tau}\textbf{b} \label{18}
\end{equation}
\begin{equation}
\textbf{b}'=-{\tau}\textbf{n} \label{19}
\end{equation}
The dynamical evolution equations in terms of time yields
\begin{equation}
\dot{\textbf{t}}=[{\kappa}'\textbf{b}-{\kappa}{\tau}\textbf{n}]
\label{20}
\end{equation}
\begin{equation}
\dot{\textbf{n}}={\kappa}\tau\textbf{t} \label{21}
\end{equation}
\begin{equation}
\dot{\textbf{b}}=-{\kappa}' \textbf{t} \label{22}
\end{equation}
along with the flow derivative
\begin{equation}
\dot{\textbf{t}}={\partial}_{t}\textbf{t}+(\vec{v}.{\nabla})\textbf{t}
\label{23}
\end{equation}
With these mathematical tools in hands now one is able to write down
the Euler equations if one expresses the velocity flow in the form
totally confined inside the torus as
\begin{equation}
\textbf{v}=v_{s}(r)\textbf{t}+v_{\theta}(r,s)\textbf{e}_{\theta}\label{24}
\end{equation}
where $v_{s}$ and $v_{\theta}$ are the toroidal and poloidal flow
components. The Euler equations yield
\begin{equation}
\frac{{\partial}_{r}p}{{\rho}_{0}}=v_{s}{{\kappa}_{0}}^{2}-v_{\theta}{\omega}_{0}+\frac{2}{r}[v_{s}-v_{\theta}]{\kappa}_{0}v_{s}\label{25}
\end{equation}
one notes that this first equation is non-linear on velocity flow
and ${\kappa}_{0}$ is the Frenet curvature. Besides one has
considered that the pressure $p=p(r)$. The remaining equations are
\begin{equation}
\frac{2}{r^{2}}v_{s}+\frac{1}{r}{v'}_{\theta}+{v"}_{\theta}={\gamma}v_{\theta}\label{26}
\end{equation}
where the dash represents the radial derivatives. The remaining
equation is
\begin{equation}
\frac{1}{r}(v_{\theta}-v_{s})+\frac{1}{r}{v'}_{s}+{v"}_{s}={\gamma}v_{s}\label{27}
\end{equation}
where the compact Riemannian operator \cite{17}
\begin{equation}
{\cal{L}}_{m}=
[\frac{d^{2}}{dr^{2}}+\frac{1}{r}\frac{d}{dr}+\frac{2}{r^{2}}]\label{28}
\end{equation}
and the self-induction or dynamo equation can be written as
\begin{equation}
{\cal{L}}_{m}\textbf{B}={\gamma}\textbf{B}\label{29}
\end{equation}
To obtain these equations a strong hypothesis simplification of
${\theta}<<1$ of small angle amplitudes is undertaken. Here one also
assumes that the poloidal rotation ${\omega}_{0}$ inside the torus
is constant and $v_{\theta}={\omega}_{0}r$. These unsteady flow
equations on the twisted torus are easily solved by assuming the
eigenvalue like expression
\begin{equation}
v_{\theta}=mv_{s}\label{30}
\end{equation}
where $m\in{\textbf{R}}$ and the terms non-linear in the Frenet
curvature were dropped. By taking the coordinate substitution
$r'=lnr$ the equations and the compact Riemannian operator can be
simplified since
\begin{equation}
\frac{d}{dr'}:=\frac{1}{r}\frac{d}{dr}\label{31}
\end{equation}
By substituting these constraints on the equations (\ref{26}) and
(\ref{28}) yields the equations
\begin{equation}
{2}v_{s}+m{v'}_{s}+m{v"}_{s}={\gamma}mv_{s}\label{32}
\end{equation}
where the dash represents the radial derivatives. The remaining
equation is
\begin{equation}
(m-1)v_{s}+{v'}_{s}+{v"}_{s}={\gamma}v_{s}\label{33}
\end{equation}
Multiplying the last equation by m and subtracting the result from
the first equation (\ref{30}) one obtains
\begin{equation}
[2-m(m-1)]v_{s}=0\label{34}
\end{equation}
This equation can be expressed in the eigenvalue matrix form
\begin{equation}
\textbf{M}_{\textbf{dyn}}=\pmatrix{(2-m(m-1))&0\cr0&1\cr}\qquad
\label{35}
\end{equation}
which yields the simple eigenvalue equation
\begin{equation}
m^{2}-m-1=0\label{36}
\end{equation}
which, in turn, yields
\begin{equation}
m_{\pm}=\frac{1\pm{\sqrt{5}}}{2}\label{37}
\end{equation}
which is similar to the above Arnold's eigenvalue for the uniform
stretching Riemann dynamo metric. Note that if one chooses $m_{+}$
the poloidal velocity is higher than the toroidal one while when
$m_{-}$ is chosen the toroidal velocity flow dominates in intensity
over the poloidal one. The last constraint on the torus flow
equations is the solenoidal vorticity $\vec{\Omega}$
\begin{equation}
{\nabla}.\vec{\Omega}=0\label{38}
\end{equation}
yields
\begin{equation}
{v'}_{s}=-\frac{1}{r}\label{39}
\end{equation}
which yields $v_{s}=-lnr$. Substitution of this expression into the
pressure equation
\begin{equation}
p(r)={\rho}_{0}[{{\omega}_{0}}^{2}r-m{\kappa}_{0}(lnr)^{2}]\label{40}
\end{equation}
Note that as $r\rightarrow{0}$ this reduces to
\begin{equation}
p(0)=-{\rho}_{0}m{\kappa}_{0}(lnr)^{2}\rightarrow{\infty}\label{41}
\end{equation}
which shows that the pressure "explodes" at the internal torus axis.
This shows that the flow is unstable inside the torus in the absence
of dissipation. One is lucky enough to understand that in
laboratories this is a very unlikely situation since as far as
liquid metals and even laminar plasmas are concerned the magnetic
and fluid Reynolds numbers are finite and as one shall see in the
next section in these cases the flow is stable inside the torus,
which turns the task of building a liquid sodium dynamo flow
detector in laboratory less difficulty.
\newpage
Let us now use the computation of the vorticity used above and given
by
\begin{equation}
\vec{\Omega}=-\frac{1}{r}[\textbf{t}cos{\theta}-\textbf{b}sec{\theta}]{\kappa}_{0}v_{s}\label{42}
\end{equation} to compute the ${\alpha}$-dynamo \cite{16} effect
\begin{equation}
{\alpha}=\vec{\Omega}.\textbf{v}\label{43}
\end{equation}
which then yields
\begin{equation}
{\alpha}=\frac{1}{r}[m-1]{{\kappa}_{0}}^{2}{v_{s}}^{2}\label{44}
\end{equation}
which shows clearly that the $\alpha$ effect is a second-order
effect in the torus curvature. Substitution of the m values, after a
straightforward and simple computation one obtains
\begin{equation}
{\alpha}=\frac{1}{r}\frac{(1\pm{\sqrt{5}})}{2}{{\kappa}_{0}}^{2}{v_{s}}^{2}\label{45}
\end{equation}
from which one notes that at the torus internal axis the $\alpha$
effect is quite strong enhancing the dynamo effect. Of course the
$\alpha$ expression is actually random and this cannot be generally
true. \section{Loops in twist torus non-dynamo maps and flows} In
this section one shall use the mathematical formalism of the Frenet
frame of the first section to build the loop twist torus dynamo
filamentary flows using the eigenvalue problem. Let us consider the
matrix of the magnetic field in the Frenet frame in the form
\begin{equation}
\textbf{B}=(B,0,0)\label{46}
\end{equation}
and the self-induction equation
\begin{equation}
d_{t}\textbf{B}=\textbf{B}.{\nabla}\textbf{v}+{\eta}{\Delta}\textbf{B}\label{47}
\end{equation}
By taking the compact Riemannian dynamo operator, without stretching
terms as
\begin{equation}
{\cal{L}}_{\eta}={\eta}{\nabla}^{2}-{\gamma}\label{48}
\end{equation}
One shall note that due to the absence of the stretch term one shall
have to constraint direction of the rotational flow vorticity in
order to have a non-dynamo, since as one shall note from the
computations the Frenet torsion shall vanish and therefore the flow
is a planar flow and since incompressibility is assumed, from the
Zeldovich anti-dynamo theorem dynamo action shall not be possible.
By making use of the Riemannian line element along the filament
\begin{equation}
{ds^{2}}_{0}={K_{0}}^{2}ds^{2} \label{49}
\end{equation}
This metric can be obtained from the torus or flux tube metric
above, by taking the filament or very thin tube approximation of
$r\approx{0}$. By Taking into account that now the gradient operator
is written as
\begin{equation}
{\nabla}=\textbf{t}{K_{0}}^{-1}{\partial}_{s}\label{50}
\end{equation}
Substitution of (\ref{46}) in the form $\textbf{B}=B\textbf{t}$ and
the gradient operator (\ref{49}) into equation (\ref{47}) leads to
the matrix form
\begin{equation}
\textbf{M}_{\textbf{fil}}-{\gamma}\textbf{I}=-{K_{0}}^{-2}{\gamma}\pmatrix{1+\frac{\eta}{\gamma}A&0\cr0&\frac{\eta}{\gamma}B+C\cr}\qquad
\label{51}
\end{equation}
Here $A:= K_{0}{\kappa}'{\kappa}$, $B=\frac{{\kappa}}{{K_{0}}^{2}}$,
$C=\frac{\kappa}{\gamma}v_{0}$. Where the $3{\times}3$ matrix
missing term $M_{33}$ equals $\frac{\eta}{\gamma}{\tau}K_{0}$ but
due to the computations this term vanishes due to the vanishing of
torsion. Here $\textbf{I}=\pmatrix{1&0\cr1&1\cr}\qquad$is the unit
matrix. Taking the determinant of the expression (\ref{51}) one
obtains the eigenvalue equation as
\begin{equation}
Det[\textbf{M}_{\textbf{fil}}-{\gamma}\textbf{I}]=0 \label{52}
\end{equation}
which implies the algebraic equation
\begin{equation}
\frac{\eta}{\gamma}(BAC)+C+(\frac{\eta}{\gamma})^{2}BA=0 \label{53}
\end{equation}
which results in
\begin{equation}
\frac{\eta}{\gamma}(B_AC)+C+(\frac{\eta}{\gamma})^{2}BA=0 \label{54}
\end{equation}
which indicates that either the dynamo is slow, since ${\gamma}$ may
vanishes when the diffusion ${\eta}$ vanishes, or the Frenet
curvature is constant. In this last case the dynamo could be fast,
however this is an unphysical solution since in turbulent flow for
example, a constant filamentary curvature should be unstable and
variations in curvatute along the filament would produce again a
slow or marginal dynamo. Actually, the vanishing of filament torsion
implies that the flow is planar and since incompressibility
condition was used throughout the computations according to
Zeldovich anti-dynamo theorem there is no dynamo action at all.
Actually this was expected due to the absence of the stretching
terms. Some maps as the Baker's maps are non-dynamo maps
\cite{18,19} such as this one one just considered here.
\newpage

\section{Conclusions} In this paper the importance of
investigation the twisted torus dynamo unsteady flows, in
non-dissipative case for the building of the torus dynamo
experiments with liquid metals was discussed. In particular it is
shown that dynamo flow solutions can be found with special
eigenvalue problem. More recent chaotic fast dynamos were obtained
by Lau-Finn result which has a growth rate of ${\gamma}=0.077$ for
$Re_{m}={\infty}$. They also obtain for $Re_{m}={1000}$ a
${\gamma}=0.076$. Growth rates as high as $3$ have been found in
Perm liquid sodium dynamo torus, as has been computed by Dobler et
al \cite{2}. The ideas developed here followed the path of realistic
dynamo maps obtained from the Gilbert's shearing generalization of
Arnold's cat map. Topology of chaos \cite{20} is a nice subject
which deals with torsion and knott theory , writhe numbers to useful
in the further investigation of the topics discussed here. In the
filamentary case discussed above actually an application of
anti-dynamo theorem of Vishik's \cite{21} has been done since slow
dynamos in non-stretching flows are obtained.
\section{Acknowledgements}: \newline Thanks go to
R. Ricca and Dmitry Sokoloff for helpful discussions on the subject
of this paper. Financial supports from Universidade do Estado do Rio
de Janeiro (UERJ) and CNPq (Brazilian Ministry of Science and
Technology) are highly appreciated.

\end{document}